\documentstyle[12pt]{article}
\begin{document}
\setlength{\oddsidemargin}{20 mm}
\setlength{\topmargin}{-10 mm}
\setlength{\textwidth}{17.0cm}
\setlength{\textheight}{23.0cm}
\renewcommand {\baselinestretch}{1.5}
\baselineskip 18 pt plus 1pt minus 1pt
\setcounter{page}{0}
\begin{titlepage}
\vspace*{-2 cm}
\begin{flushright} May 11, 1993\\
CEBAF-TH-93-07,\\
WM-93-105\end{flushright}\vspace*{0.3 cm}
\begin{center}
PRODUCTION, COLLECTION AND UTILIZATION\\
OF VERY LONG-LIVED HEAVY CHARGED LEPTONS\vspace{1.0cm}\\
J. L.  Goity \vspace*{3mm}\\
 {\it  Continuous Electron Beam Accelerator Facility\\
        Newport News, VA 23606, USA.}\vspace*{0.6 cm}\\
W. J. Kossler and Marc Sher\vspace*{3mm}\\
{\it Department of Physics, College of William and Mary, \\
Williamsburg, VA 23187}, USA.\vspace*{1 cm}\\
\end{center}
\begin{abstract}
If a fourth generation of leptons exists, both the neutrino and its
charged partner must be heavier than 45 GeV. We suppose that the neutrino
is the heavier of the two, and that a global or discrete symmetry
prohibits intergenerational mixing. In that case, non-renormalizable
Planck scale interactions will induce a very small mixing; dimension five
interactions will lead
  to a lifetime for the heavy charged lepton of
  $O(1-100)$ years. Production of such particles is discussed, and
it is shown that a few  thousands can be produced and collected at
a linear collider. The possible uses of these heavy  leptons is also
briefly discussed.
\end{abstract}
\end{titlepage}
\newpage
\setcounter{page}{1}

It has now been established $\cite{lep}$ at LEP that there are three light
neutrinos.
If a fourth generation exists, the mass of both its neutrino, N,  and the
associated charged lepton, L, must exceed 45 GeV or it would have
been observed
at LEP.  Such a generation would be
unique in that the neutrino and charged lepton will have masses which do not
differ by much more than an order of magnitude.

In this Brief Report, we consider the implications of a fourth generation
with the following two properties:  (a) the mass of the neutrino is greater
than the mass of the charged lepton and
(b) some symmetry (discrete or global),
which remains unbroken as   electroweak symmetry
breaking takes place, prevents intergenerational mixing (in the lepton
sector).
  Neither of these
assumptions is particularly implausible.  The first simply chooses a
particular
half of the allowed parameter space.  The second is just an extension of the
familiar electron-number, muon-number and tau-number conservation laws.

As a consequence of these two properties, the charged lepton, L, would appear
to
 be absolutely stable.  This is a cosmological disaster.  The abundance of
these
leptons today would be large enough that they could not have escaped detection
in terrestrial experiments (searches for heavy hydrogen in water).  A detailed
analysis of the cosmological bounds can be found in Ref. $\cite{yao}$,
 where an upper
bound of roughly 100 years is found on the lifetime of a charged lepton.  For
lifetimes   greater than 100 years, the photons emitted in the decay
would distort the microwave background radiation more than is observed by
COBE.

The model is not necessarily excluded, however.  It has long been recognized
 that black holes violate global and discrete symmetries $\cite{hawk}$,
and one would
expect quantum gravitational effects to also violate such symmetries (a
discrete
gauge symmetry $\cite{wilczek}$, however, remains unspoiled by such
effects).  Thus,
higher-dimensional operators, scaled by the Planck mass, which violate such
symmetries can be included.  In the case of baryon number conservation,
dimension six operators will lead to proton decay with a lifetime of
$O(10^{45})$ years.  More recently, it has been realized
$\cite{axion}$ that axion models are ruled out if, as one expects, quantum
gravitational effects do not respect the Peccei-Quinn symmetry.  In our case,
one would expect higher-dimensional operators to violate the symmetry which
prevents intergenerational mixing, leading to a finite, albeit very long,
lifetime for the L. We will not address  here
the origin of the symmetries responsible for L being nearly stable.

How long does one expect the lifetime to be?  If the mixing angle is
 $\theta$,
then the decay rate is ${G_Fm_L^3\sin^2\theta/ (8\pi\sqrt{2})}$, or
approximately  $10^{24}\sin^2\theta\ {\rm sec}^{-1}$ for a lepton of
 mass $200$
GeV.  If the lowest-dimension Planck scale operator which violates
the symmetry
is dimension six (or higher), then the mixing angle will be smaller than
$M^2_W/M^2_{Pl}$, leading to a lifetime in excess of $10^{37}$ years,
 which is
cosmologically unacceptable.  However, if a dimension five operator
violates
the symmetry, then the mixing angle will be $O(M_W/M_{Pl})$, leading
 to a
lifetime of approximately $1-100$ years,  close to the cosmological
 bound.

Two models will illustrate this point.  Suppose the $L$ has a mass of
 $250$ GeV, and consider
 mixing between the L and the $\tau$.  In one model, one can introduce
a gauge
singlet scalar\footnote{A natural choice
for such a singlet in a supersymmetric model would be the fourth generation
scalar neutrino; if it gets a vacuum value, then this would naturally
explain why the fourth generation neutrino is heavy.}, $S$, and write a
dimension five operator:
\begin{equation}
 \frac{f_{L\tau}}{M_{Pl}}\;\bar{L}_{L} \Phi \tau_{R} \;  S +
\frac{f_{\tau L}}{M_{Pl}}\;\bar{\tau}_{L} \Phi L_{R}
 \; S +{\rm\ h.c.},
\end{equation}
where  we have assumed that $L_{L}$ is a doublet under
$SU(2)_{L}$,
 $\Phi$ is the standard model Higgs boson and $f_{L\tau}$
and $f_{\tau L}$ are  constants of $O(1)$.
In another model, one can have the $L$ being a mirror fermion (left-handed
singlet and right-handed doublet)$\cite{roos}$, and write, in obvious
notation,
 the dimension five operators:
\begin{equation}
 \frac{f_{L\tau }}{M_{Pl}}\; \bar{L}_{L} \Phi^{\dagger} \Phi\tau_{R} +
\frac{f_{ \tau L}}{M_{Pl}}\;\bar{ \tau}_{L} \Phi\Phi^{\dagger} L_{R}
  +{\rm\ h.c.},
\end{equation}
  When $\Phi$
(and in the first model, $S$) acquire vacuum expectation values, mixing
between
the $L$ and the $\tau$ will be induced.  In the latter model, the lifetime
turns
out to be
$\sim {10\over f^2}\left({250 {\rm GeV}\over m_L}\right) {\rm years}$, with
$f$ of $O(1)$;
 in
the former model this is multiplied by
$ (250 \, {\rm GeV}/\langle S \rangle)^{2} $.
 While we are
not advocating any particular model, one can see that plausible models with a
lifetime in the $1-100$ year range can easily be constructed.

Such a long lifetime  leads to the possibility that one could produce
these particles at a high energy collider, stop them in some material,
physically transport them away from the  detector environment and study them
at
length.  We now consider these possibilities.

A particle with several tens or hundreds of GeV of kinetic energy cannot
 be stopped, and thus one would want to produce the $L$'s at an
electron-positron collider just above threshold. In the case that the heavy
fermion family  has the same electroweak couplings as the
lighter lepton families, the production cross-section
is given by:
\begin{equation}
\frac{d\sigma}{ d\cos\theta}=\frac{\beta}{32\pi s}\,\left[\,\xi_1\,
(1+\cos^2\theta)+
(1-\beta^2)\,\xi_2\,\sin^2\theta+\xi_{F-B}\,\cos\theta\,\right],
\end{equation}
where $\theta$ is the   angle, in the c.m. frame,
 between the outgoing $\rm L^{-}$ and
the incoming electron, and
\begin{eqnarray}
\xi_2&=&(e^2+g_z^2)^2+a^2\;\;g_z^4 \nonumber \\
 \xi_1&=&\xi_2+\beta^2\; a^2\;\;g_z^4\;\;(1+a^2) \\
 \xi_{F-B}&=&-\,8\;a^2\;\beta \;g_z^2\;\;(g_z^2+\frac{1}{ 2}\;e^2)\nonumber.
\end{eqnarray}
Here we have defined $a\equiv 1/(4\sin^2\theta_W-1)$,
 $e \equiv g\sin\theta_W $, and $ g_z\equiv\
(g/4a)\,\sec\theta_W $.  The result is given in the limit
of $\sqrt{s}>>M_Z$; the $M_Z$ dependence can be included by multiplying
$g^2_z$
by $s/(s-M^2_Z)$.   The kinetic
energy of the $L$, of course, is ${1\over 2}M\beta^2$.  Note that in the limit
of small $\beta$, the distribution is isotropic.  If one has a luminosity of
$3\times 10^{33}\ {\rm cm}^{-2}{\rm sec}^{-1}$ (which is the expected
luminosity of the Next Linear Collider), then in a year of running the number
of
$L$'s produced is approximately $7\times 10^4\beta$.  In Table 1, the
precise number of $L$'s produced as a function of $\beta$ is given (taking the
$L$ mass to be $200$ GeV), as well as the kinetic energy and stopping distance
in
liquid argon. We see that many thousands of sufficiently low energy $L$'s can
be
produced.  For definiteness, we will take $\beta=0.2$,
and thus $14,500$ can be
produced annually.

Many properties of the $L$ will be determined directly from production
cross-section.  The mass can be measured precisely from the energy
threshold, and the spin will be immediately determined from the angular
distribution.  Other quantum numbers can be  determined from the
forward-backward asymmetry, given by
\begin{equation}
 A_{F-B}=\xi_{F-B}/(\frac{8}{3}\;\xi_1+(1-\beta^2)\;\xi_2).
\end{equation}
This asymmetry vanishes at $\beta=0$ and, for $\beta\leq 0.6$, increases
roughly linearly with a slope of $\sim 0.4$.  For $\beta=0.2$, the asymmetry
of $0.08$ could be easily detected with an integrated luminosity of $10^{41}
{\rm cm}^{-2}$; we estimate that an asymmetry of $0.01$ could be measured in
this case.  Measurement of this asymmetry offers the possibility of
determining whether $L$ belongs to a doublet of $SU(2)_L$ by checking that the
axial coupling of the $L$ to the $Z$ is the same as that of the electron.
We estimate that $g_A$ can be measured to an accuracy of ten percent; $g_V$
is naturally small and thus cannot be determined accurately. If the heavy
lepton
belongs to a mirror lepton family, all the above applies except that now
the forward-backward asymmetry has the same
 magnitude but the opposite sign  $\cite{banks}$.
  Finally, one can
consider detecting a possible electric dipole moment (EDM) of the $L$.  In
many
models, the EDM rises as the cube of the lepton mass,
 and could be very large in
this case.  However, by considering the effects of an EDM on the angular
distribution (see Ref. $\cite{delaguila}$ for a detailed expression), we can
see that
the above luminosity will only allow an EDM greater than $2\times 10^{16}$
e-cm.
to be detected. This value is at least an order of magnitude greater than the
largest values expected theoretically $\cite{soni}$, and thus it is unlikely
to be seen.

Of greater interest, of course, is the possibility of capturing the $L$'s.
The stopping distance in liquid argon (Table I) is certainly experimentally
tractable; the stopping distance in other substances will  scale
roughly inversely with the density, and thus will not differ greatly.
The charge could be determined by observing the curvature of the track in a
magnetic field (a $2$ Tesla B-field will bend a lepton of mass $200$ GeV  with
$\beta=0.2$ with a radius of curvature of $70$ m.); this determination is
essential in obtaining the forward-backward asymmetry mentioned in the above
paragraph.

$L^-$ will most likely be captured by the highest $Z$ nucleus in the stopping
medium.  The result is a heavy nucleus of charge $Z-1$.  Thus, if the stopping
medium were argon, the result is a chlorine-like atom.  This could then be
chemically separated from the inert argon.  Alternatively, stopping $L^-$ in
Na
or K would lead to an inert Ne or Ar like atom which could be boiled off and
then condensed onto a collector.

The $L^+$ will probably pick up an electron upon stopping.  Stopping in an
inert gas might delay this capture process, but the lepton is unlikely to
remain charged for times comparable to the collection and extraction times.
For example, positive muons can be extracted with an efficiency of $10^{-4}$
from argon in a microsecond, however, there is already evidence that electrons
diffuse toward the muon in this amount of time.  Thus, the $L^+$ will almost
certainly be neutralized.  The $(L^+e)$ atom produced in this way will be
chemically identical to hydrogen, and equally chemically active.  Thus, for
example, one could stop the $(L^+e)$ atom in a chlorinated liquid and then it
would react with the Cl to produce $LCl$ which could then be evaporated and
collected.  To avoid missing half of the $L$'s, one might wish to use the same
stopping medium, such as argon, for each, in which case one would have to
chemically extract the hydrogen-like atom from argon.

Since the mass, energy and charge of the $L$'s will be known before they
enter the stopping medium, the range will be known to within a few
centimeters.
With a segmented stopping medium, one could isolate their location to within
a volume of a few tens of milliliters.  Since the  production rate of Table I
corresponds to roughly one event per hour, one could remove that volume and
perform the chemical extraction at another site.  One might even avoid
chemical
extraction altogether and use a mass spectrometer to isolate the $L$'s (since
their mass is known); more likely both would be used.  In any event,
collection
of thousands of $L^+$'s and $L^-$'s should be possible.

 What can one do with a few thousand $L$'s?  The first important
measurement would be the lifetime and decay products (it presumably decays
into
a $\nu$ and a real or virtual $W$).  The $L$'s could be placed in
 an underground
detector such
 as super-Kamiokande, and the decays could be
individually measured.  Since the cosmological bound on the lifetime is less
than
$100$ years or so, the lifetime could be determined fairly rapidly.

A few thousand $L$'s would be  useless as an  energy storage
device ($10^4\, M_L\,c^2$ is
less than a millijoule).  Since the reduced mass of an $(L^-p)$  hydrogen atom
is
the proton mass, the $L$'s would catalyze D-D and D-T  fusion
much as does a $\mu^{-}$.   However, at one
fusion per nanosecond (which is the rate at which muons catalyze fusion--the
$L$'s should be much slower due to their greater mass), each $L$ will produce
roughly a milliwatt, and thus a few thousand would not give a practical energy
source $\cite{okun}$.

Finally, one might be able to use the $L$'s to study nuclear structure.  The
X-rays emitted when an $L^-$ cascades down towards the center of a nucleus
would give information about the electromagnetic structure of the nucleus (it
is doubtful, however,  whether more information can be obtained in this manner
than by electron scattering, such as at CEBAF).  A detailed discussion of the
possible uses of heavy stable particles can be found in Ref. $\cite{zweig}$,
where
Zweig discusses the effects of very heavy stable quarks. He shows that large
nuclei would fission upon capture of a heavy stable particle, that alpha decay
could be facilitated upon capture of an $L$ (for example, introducing a
stable quark of mass $300$ MeV and charge $-4/3$  into thorium will reduce its
half-life by a factor of $10^{36}$),
and that a new class of  molecules would be
produced.  Unfortunately, many of the effects he considers would require many
more than a few tens of thousands of $L$'s in order to be of practical
use $\cite{okun}$.

In this work we have considered the production, collection and utilization of
very long-lived heavy leptons.  It must be emphasized that our only
assumptions
are that (a) a fourth generation of fermions exists, (b) the neutrino of that
generation is heavier than the charged lepton and (c) that a discrete or
global symmetry prohibits intergenerational mixing.  Given the first
assumption, we do not believe that either of the other two is particularly
unlikely.  It has been shown that the lifetime can be of the order of
 ten years, and
that thousands of these heavy leptons could be collected at an
electron-positron collider with an energy slightly above threshold.  Although
we have speculated on the possible uses of such leptons, such speculation is
undoubtedly premature--clearly there will be applications that we have not yet
imagined. More important, the discovery of a heavy lepton
with the properties considered here  should have a deep
significance for a further understanding of the generation problem.

We thank Bob Welsh for useful discussions.  This work was supported in part by
the National Science Foundation

\newpage

\vspace*{1cm}

{\Large{\bf Table 1}}\\

\vspace*{2cm}
{\large
\begin{tabular}{|c|c|c|c|} \cline{1-4}
$\beta$ & 0.3 & 0.2 & 0.1 \\
\cline{1-4}
$\sigma(pb)$&  0.24 &  0.16 &   0.08 \\ \cline{1-4}
{\rm No./yr.}& 22,000 & 14,500 & 7,600 \\ \cline{1-4}
{\rm K.E.(GeV)} &  9.0 &  4.0 &  1.0 \\ \cline{1-4}
d\ {\rm in\  liquid\ Ar}\ ({\rm cm.}) &  225 &  55 &  5 \\ \cline{1-4}
\end{tabular}}
\vspace*{5cm}
\newpage
\vspace*{1cm}

{\Large{\bf Table Caption}}\\
\vspace*{2cm}\\

{\bf Table 1}: Production cross section, number of heavy leptons produced per
year,
 kinetic energy and stopping distance in liquid Ar as a function of $\beta$
for
a 200 GeV charged heavy lepton.
\vspace*{5cm}
\end{document}